# Refractive-index-matched hydrogel materials for measuring flow-structure interactions

Margaret L. Byron and Evan A. Variano

University of California – Berkeley

Department of Civil and Environmental Engineering

**ABSTRACT**

In imaging-based studies of flow around solid objects, it is useful to have materials that are refractive-index-matched to the surrounding fluid. However, materials currently in use are usually rigid and matched to liquids that are either expensive or highly viscous. This does not allow for measurements at high Reynolds number, nor accurate modeling of flexible structures. This work explores the use of two hydrogels (agarose and polyacrylamide) as refractive-index-matched models in water. These hydrogels are inexpensive, can be cast into desired shapes, and have flexibility that can be tuned to match biological materials. The use of water as the fluid phase allows this method to be implemented immediately in many experimental facilities and permits investigation of high Reynolds number phenomena. We explain fabrication methods and present a summary of the physical and optical properties of both gels, and then show measurements demonstrating the use of hydrogel models in quantitative imaging.

Address for correspondence: mbyron@berkeley.edu, 202 O'Brien Hall, Berkeley, California 94720-1712

# 1 Introduction

Particle Image Velocimetry (PIV) has been an important tool for measurements in fluid mechanics (Raffel et al 2007). However, this technique is not easily extended to the specific case of flow around solid objects. This is because PIV requires the use of a laser light sheet to illuminate the region of interest. When measuring flow around an opaque or translucent object, the object itself will interfere with the illumination by casting shadows and/or scattering light. Of specific interest to our research is the case of "macroparticles" (regular and irregular three-dimensional shapes, of lengthscale $\approx$ 1cm) suspended in a flow. When these macroparticles cast shadows, we often cannot see their wakes, nor can we see into the interior of a dense suspension. These problems are also encountered when using PIV around models (e.g., of organisms or turbomachinery): shadows eliminate large portions of the image area, preventing a complete analysis of the flowfield (Schiacchitano et al 2012).

To resolve this issue, refractive index matching (RIM) has been employed with great success in fluid measurement techniques such as Laser Doppler Velocimetry (Budwig 1994). This strategy is easily extended to PIV (Hassan and Dominguez-Ontiveros 2008). In RIM-PIV, the test objects (in our case, macroparticles) are made of a material that is transparent and refractively matched to the surrounding fluid. This avoids blockage or distortion of the laser light sheet, and grants optical access to the entire flowfield. One commonly used refractive-index-matched pair is mineral oil and glass or fused quartz (Stoots et al 2001; Thompson et al 1987; Ezzein and Bathurst 2011). This pair has been used by biologists to conduct dynamically-matched experiments of low-Reynolds-number phenomena such as lobster antennule flicking (Reidenbach et al 2008).

It has also been used by hydrologists to study flow through porous media (Lachhab et al 2008). The high viscosity of mineral oil, however, precludes examination of high-Reynolds-number phenomena, including turbulent flows. By using aqueous sodium iodide solution matched with glass or acrylic, one can reach much higher Reynolds numbers, but at significant economic expense (Uzol et al 2002). Butscher et al (2012) used resin paired with anisole to investigate flow through porous structures. Neither glass, acrylic, nor resin, however, allows for the study of flexible or deformable materials, which are common in biology.

Herein we explore the use of two hydrogels, which by virtue of their chemistry are nearly refractive-index-matched to water. These hydrogels can be easily manufactured in the laboratory using injection molding (as discussed in following sections); more complex shapes can be obtained through stereolithography (Arcaute et al 2011). The material cost is small, and the use of water as the working fluid greatly expands experimental options. Furthermore, because hydrogels have adjustable density and flexibility, they can be used to model myriad objects that are of interest in biological fluid dynamics.

## 2  Materials and methods

We focus our efforts on the polyacrylamide (PAC) hydrogel. PAC is an organic polymer with subunit formula -$CH_2$CHCON$H_2$- (acrylamide). Aqueous acrylamide at low concentrations can be chemically or photochemically polymerized to form a highly water-retentive hydrogel. This gel is commonly used as a matrix for DNA gel electrophoresis, a staple technique in molecular biology. Additionally, PAC is used in agricultural soil treatment, (Sojka et al 1998, Boatright et al 1997), aesthetic surgery (Pahlua and Wolter 2010), and soft contact lenses (Christensen et al 2003).

PAC gel has a number of qualities desired for RIM-PIV: it is straightforward to manufacture, it can be cast easily into a variety of shapes, it is nearly transparent, and has a refractive index close to that of water. Its optical properties, density, and elastic modulus can be controlled by chemical composition and manufacturing method. Additionally, it can be "seeded" with internal tracers to facilitate tracking during imaging studies. Using this last feature, we will demonstrate that images of PAC macroparticles can yield detailed vector fields describing the solid-body motion of macroparticles as well as the surrounding flow field, using a standard commercial PIV system.

We also discuss briefly a second option for RIM-PIV: agarose hydrogel. Agarose is a polymer with subunit formula -$C_{24}H_{28}O_{10}(OH)_8$- (agarobiose). It is most commonly used as a laboratory growth medium for microorganisms. We have already used agarose successfully in RIM-PIV (Bellani et al 2012). Agarose is more fragile than PAC, and is less transparent, leading to excessive light scattering that disrupts imaging. Because PAC has superior optical clarity, robustness, and longevity, we prefer to use this in our experiments, and make it the main focus of this paper.

2.1   Procedure for fabricating polyacrylamide models

Plastic molds are 3D printed (purchased from Protocafe Inc. using ProJet 30000 HD printer and proprietary acrylic polymer) to create desired shapes. As an example, we discuss macroparticles shaped as prolate ellipsoids of minor axis 8mm and major axis 16mm. The molds are lightly coated with mineral oil, clamped into their assembled form, and set aside before mixing the acrylamide solution.

De-ionized water is seeded with 11μm glass spheres (Sphericel, manufactured by Potters Industries) at a concentration of 0.05% by mass, to act as embedded optical tracers for PIV. 30% acrylamide solution (manufactured by Bio-Rad Laboratories) is diluted such that acrylamide composes 8% of the total solution volume (see Table 1). The ratio of acrylamide isomers (in this case, 37.5 parts acrylamide to 1 part bis-acrylamide) is responsible for the cross-link density and therefore the consistency of the gel.[1]

A 10% aqueous Ammonium Persulfate (APS) solution is mixed using de-ionized water and ammonium persulfate crystals (manufactured by Bio-Rad Laboratories). The 10% APS solution is added to the acrylamide-seed solution, constituting 0.5% of the total solution volume. The catalyst tetramethylethylenediamine (TEMED, manufactured by Bio-Rad Laboratories) is then added to the solution in a quantity constituting 0.1% of the total solution volume. The total solution is mixed well to ensure even distribution of the two polymerizing agents (APS and TEMED) and to improve suspension of tracer particles. Excessive stirring is avoided, as oxygen will absorb the free radicals necessary for polymerization. The solution is then injected into the custom-shaped molds using a hypodermic syringe. The solution begins to polymerize into a gel in approximately ten minutes, and polymerization is complete in two to three hours.

---

[1] We note that acrylamide in its unpolymerized form is a potent neurotoxin; to avoid adverse health effects, latex gloves are worn at all times and care is taken to avoid spillage.

After complete polymerization has occurred, the macroparticles are removed from the molds and placed in containers of de-ionized water. After 24 hours in water, the macroparticles expand by approximately 10% in length dimensions. Macroparticles have been observed during 6 months of aging with no visible signs of degradation in shape or optical clarity. We store macroparticles in a screw-top container, refrigerated and submerged in a water bath[2].

**Table 1: Solute masses/volumes for three different formulations of PAC hydrogel**

|  | **8% Gel** | **12% Gel** | **16% Gel** |
|---|---|---|---|
| **30% Acrylamide/Bis** | 26.7 mL | 40.0 mL | 53.3 mL |
| **10% APS** | 0.5 mL | 0.5 mL | 0.5 mL |
| **TEMED** | 0.1 mL | 0.1 mL | 0.1 mL |
| **Sphericel** | 0.05 g | 0.05 g | 0. 05 g |
| **dH$_2$O** | 72.7 mL | 59.4 mL | 46.1 mL |
| *Total Solution Volume:* | 100 mL | 100 mL | 100 mL |

As previously discussed, a second option for RIM-PIV is agarose hydrogel. Our materials and procedure for manufacturing agarose macroparticles are as follows.

2.2   Procedure for fabricating agarose models

A beaker of de-ionized water is heated in a hot water bath to a temperature of approximately 50C. Agarose powder (Apex BioResearch Products General Purpose

---

[2] To prevent microbial growth in the water bath, a small amount of detergent is added. There is no microbial growth within the PAC macroparticles themselves.

Agarose, low electroendosmosis value) is added at 0.4% by mass (see Table 2). 13-44μm glass spheres[3] are added at 0.2% by mass (see Table 2), again to provide embedded optical tracers. The mixture is stirred until agarose powder dissolves. This mixture is injected into clamped, oiled molds as in the PAC procedure outlined above, and refrigerated for ten minutes in order to set the gel. Gelled macroparticles are removed and placed in water, as above. Agarose does not visibly expand or further hydrate after casting. Agarose macroparticles are significantly more fragile than PAC particles and must be handled with care to avoid breakage.

Degradation of agarose macroparticles by microbial life is evident after 4-6 weeks in 4°C refrigerated storage. Degradation is greatly slowed (but not stopped) by refrigeration.

**Table 2: Solute masses/volumes used in 0.4% agarose hydrogel formulation**

| Agarose powder | 2.0 g |
|---|---|
| dH$_2$O | 500 mL |
| 13-44μm tracer particles | 1.0 g |

---

[3] Larger glass spheres are used for agarose macroparticles than PAC macroparticles to accommodate differences in background light-scattering in the two gels. This is explained in detail in section 3.2.

## 3 Assessment of method

### 3.1 Physical Properties

Physical properties for the two hydrogel materials are given in Table 3. For PAC, we focus on the 8% PAC formulation[4]. All refractive indices and densities are measured at 23°C. Index of refraction is measured without embedded tracers using a refractometer (Atago Inc.) which measures a refractive index of 1.332 for the deionized water with which the particles are manufactured. Density measurements are made using Archimedes' method, again using the deionized water with which the particles are manufactured.

Hydrogels are much less rigid than other common refractive-matched materials, and this makes them potentially useful in modeling the interaction of fluid flow and flexible structures such as biological tissue. In such studies, the formulation of hydrogel should be tuned to give the desired material properties. To provide a starting point for these modifications, we measure the maximum compressive stress and strain for 8% PAC and agarose gels. Both materials exhibit non-Hookean behavior and neither exhibit complete elastic recovery. Thus a single elastic modulus cannot be defined, but the ranges shown in Table 3 may be used when rough comparisons are needed. The maximum compressive stresses and strains of 8% PAC and of Agarose are listed in Table 3; these values are in agreement with our qualitative observations of the macroparticles' relative fragility.

---

[4] Note: higher percentage formulations yield a less flexible and more dense gel, and lower percentage formulations yield an extremely flexible gel that is slightly less dense. These formulations may be useful for other researchers, but for our application (free-floating, nearly-neutrally-buoyant macroparticles) we prefer 8% PAC.

**Table 3: Properties of PAC and Agarose hydrogels**

*(All error ranges shown are the standard error)*

| Property | 8% PAC | 0.4% Agarose |
|---|---|---|
| Index of refraction | 1.349 | 1.3329 |
| Compressive breaking stress (kPa) | 36.6 ± 2.5 | 5.1 ± 0.2 |
| Compressive breaking strain (mm/mm) | 0.61 ± 0.03 | 0.36 ± 0.03 |
| Elastic modulus at low strains (near zero) (kPa) | 10.9 ± 2.2 | 1.9 ± 0.2 |
| Elastic modulus at high strains (near breaking) (kPa) | 134.0 ± 1.2 | 22.4 ± 1.6 |
| Density (g/cm$^3$) | 1.024 ± 0.003 | 1.007 ± 0.001 |
| Specific gravity | 1.026 ± 0.008 | 1.009 ± 0.006 |

3.2 Performance in PIV

The example images in Figure 1 are collected by threading an 8% PAC macroparticle onto a piece of string and suspending it in a PIV measurement plane. The surrounding flow is homogeneous and isotropic turbulence (Bellani et al 2012). The 2D vector fields shown in figure 1 were computed by tracking tracer particles either within the moving macroparticle (Figure 1a) or suspended in the surrounding flow (Figure 1b). The resulting data show that it is possible to track PAC macroparticle translation using common PIV technology (we used commercial software from LaVision inc.). Gradients in the within-particle velocity field allow us to calculate particle rotation, which is useful in its own right and to improve the precision of particle translation measurements (Bellani and Variano 2012). The macroparticle scatters slightly more light than the water phase, allowing the two to be separated via basic image processing. This technique has also been used with agarose to describe the interactions between macroparticles and homogeneous, isotropic turbulence (Bellani et al 2012). In this case, the light scattering by agarose made these macroparticles far more challenging to use than PAC particles.

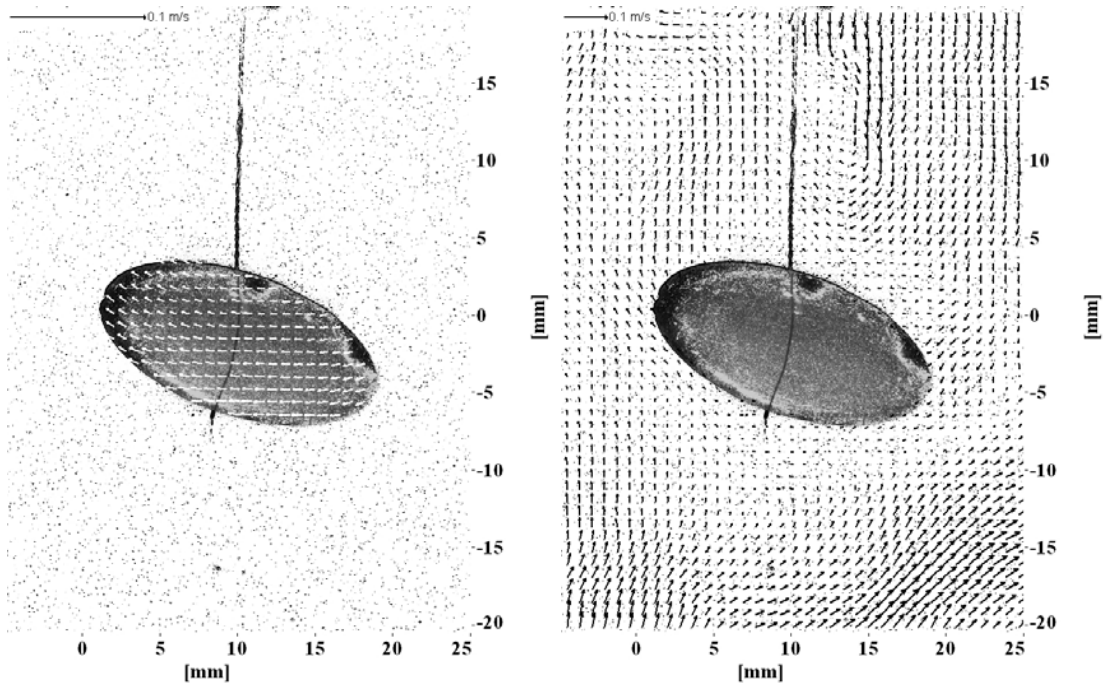

**Figure 1: PIV showing (a) solid-body motion of points within moving macroparticle and (b) turbulent flow in surrounding water**

3.3  Limitations

Two major limitations of this technique are that (1) injection molding becomes challenging for features smaller than ≈ 1 mm and that (2) hydrogels are inherently flexible, thus attempts to formulate extremely rigid particles may interfere with their optical properties. However, in cases where rigidity is required, there are many choices for refractive-matched pairs (see section 1).  Currently, model size is limited by manufacturing method (not by the material itself); future manufacturing methods may improve upon those stated here.

As previously mentioned, hydrogel particles are not completely transparent, and scatter light.  Some degree of light scattering is useful in most applications (such as the one illustrated by Figure 1), otherwise the gel becomes essentially invisible.  We have found that the optimal amount of scattering depends on the macroparticle shape, camera setup, and illumination.  Thus, some adjustment will be needeed in any new application.

Lastly, if very large quantities of particles or models are desired, a less time- and labor-intensive approach may be warranted.  Again, this limitation is that of the manufacturing method and not of the material itself.

**4   Discussion**

Both PAC and agarose, as we have produced them, are near neutral buoyancy (within 3% of the density of water).  Both also have indices of refraction that are within 2% of water.  PAC is robust to mechanical stresses, with a high maximum compressive strain.  This makes it useful for flows in which macroparticles experience a high degree of stress, e.g. in turbulence.  Agarose is less resilient than PAC, and degrades readily over time due to microbial activity. However, it can still be used in RIM-PIV experiments (e.g. Bellani et al 2012).  In its favor, agarose has less health risk during production, is less expensive, and can be more quickly mass-produced.

The potential applications of this technique are wide-ranging—relevant to biology, engineering, medicine, and many other fields. In biomechanical studies, hydrogel models of anemones, fish, or macroplankton could provide new insight into functional morphology and biohydrodynamics. Environmental engineers could also investigate sedimentation processes with this technique. The adjustable flexibility and deformability of hydrogels points towards the modeling of tissues, including flow through blood vessels or porous structures. We have illustrated only one application here, but we are confident that this method can be tuned and used for a wide variety of functions.

## 5 Conclusion

We have developed two new possibilities for refractive index matching in particle image velocimetry: polyacrylamide and agarose hydrogels. These materials are refractively matched to water, allowing measurement of high-Reynolds number flow surrounding the hydrogel. Fabrication is simple and low-cost, and internal tracers can be integrated into the gel to reveal the object motion in great detail. In addition, the flexibility and compressibility of hydrogels also makes them ideal for studying flow around modeled biological tissues, unlike previous RIM-PIV materials. We have demonstrated the materials' utility by presenting PIV vector fields from both the interior and exterior of hydrogel macroparticles. This technique provides many new opportunities for the investigation of flow around flexible objects.

## ACKNOWLEDGEMENTS

The authors would like to thank Gabriele Bellani, Audric Collignon, and Colin Meyer for their contributions to the development of this method in agarose hydrogels, Michael Nole for assisting in image collection, and Wei-Qin Zhuang for his invaluable assistance with the adaptation of polyacrylamide techniques. Testing facilities were provided by the Department of Civil and Enviromental Engineering and the Center for Integrative Biomechanics in Education and Research (CiBER). This work was supported by NSF IGERT #0903711.